\documentclass[10pt]{article}
\usepackage{epsfig}
\usepackage{a4,isolatin1}
\usepackage{amsmath,amsfonts,latexsym, amssymb}
\usepackage{theorem}
\theorembodyfont{\upshape}
\newtheorem{satz}{Theorem}[section]
\newtheorem{defi}[satz]{Definition}

\newtheorem{lemma}[satz]{Lemma}
\newtheorem{koro}[satz]{Corollary}

\newtheorem{conclusion}[satz]{Conclusion}
\newtheorem{ob}[satz]{Observation}

\newtheorem{prop}[satz]{Proposition}

\newcommand{\tit}{\textit}

\newcommand{\Z}{\mathbb{Z}}

\newcommand{\fo}{\overline{f}}
\newcommand{\fu}{\underline{f}}
\newcommand{\go}{\overline{g}}
\newcommand{\gu}{\underline{g}}
\newcommand{\diam}{\textnormal{diam}}
\newcommand{\dime}{\textnormal{dim}}

\begin{document}
\thispagestyle{empty}
\begin{center}
\vspace*{1.0cm}
\setlength{\baselineskip}{1.6em}
{\LARGE{\bf An Analysis of the Transition Zone\\ between
 the various Scaling Regimes\\ in the Small-World Model\\}
}

\vspace{1 cm}

{\large {\bf Andreas Lochmann\quad  Manfred Requardt}}

\vspace{0.2 cm}

Institut f\"ur Theoretische Physik, Universit\"at G\"ottingen, \\
Tammannstra{\ss}e 1, 37077 G\"ottingen, Germany \\
(E-mail: requardt@theorie.physik.uni-goettingen.de)

\end{center}

\vspace{0.5 cm}

\begin{abstract}
  We analyse the so-called small-world network model (originally
  devised by Strogatz and Watts), treating it, among other things, as
  a case study of non-linear coupled difference or differential
  equations. We derive a system of evolution equations containing more
  of the previously neglected (possibly relevant) non-linear terms. As
  an exact solution of this entangled system of equations is out of
  question we develop a (as we think, promising) method of enclosing
  the ``exact'' solutions for the expected quantities by upper and
  lower bounds, which represent solutions of a slightly simpler system
  of differential equation. Furthermore we discuss the relation
  between difference and differential equations and scrutinize the
  limits of the spreading idea for random graphs. We then show that
  there exists in fact a ``broad'' (with respect to scaling exponents)
  crossover zone, smoothly interpolating between linear and
  logarithmic scaling of the diameter or average distance. We are able
  to corroborate earlier findings in certain regions of phase or
  parameter space (as e.g. the finite size scaling ansatz) but find
  also deviations for other choices of the parameters. Our analysis is
  supplemented by a variety of numerical calculations, which, among other
  things, quantify the effect of various approximations being made.
  With the help of our analytical results we manage to
  calculate another important network characteristic, the (fractal)
  dimension, and provide numerical values for the case of the
  small-world network.\\[0.3cm]
Catchwords: Small-World Networks, Non-linear Difference Equations

\end{abstract} \newpage
\section{Introduction}
As part of a broader interest in complex systems, the analysis of
large networks of interacting agents or simply certain degrees of
freedom is currently under intense study. Recently a presumably
far-reaching core-concept came to the fore, called the
\tit{small-world effect}, (for an incomplete list of references see,
for example \cite{SS3} to \cite{Weigt}). To put it briefly, the
presence of a surprisingly small number of random edges, inserted in
an initially quite regular graph, may have drastic effects as to the
average distance between nodes or the expected diameter of the
network. These additional random edges, called short cuts, may
typically connect regions which have been quite a distance appart in
the original regular graph, thus effecting a drastic shrinkage of
average distance or diameter in certain regions of parameter space.

It is perhaps noteworthy that we detected a similar phenomenon in
quite a different area of modern physics (quantum space-time physics)
at almost the same time, being completely unaware of similar findings
in other fields of natural science. We called this phenomenon a
\tit{microscopic wormhole structure} (\cite{RequRen},\cite{Requ1}).

To understand this smallworld effect in more quantitative terms, a simple
model, the so-called \tit{Strogatz-Watts-model}, was investigated in
more detail in \cite{Amaral1},\cite{Newman1},\cite{Newman2} and a
little bit later also in \cite{Barbour}.

In its most tractable form it consists of $N$ linearly ordered
vertices (nodes) with periodic boundary condition (i.e. node $x_N$ is
linked to node $x_1$). In general each node may also be linked to its
regular neighbors up to order $k$. The generic case is already
given for $k=1$ (i.e. nearest neighbors only or $\Z_N$).

To mimic the random-rewiring of edges of the original
Strogatz-Watts-model, it is convenient to superimpose the given
regular graph by a random graph, living over the same set of vertices.
While we prefer to introduce the so-called \tit{edge-probability} $p$,
that is, the independent probability for the existence of an edge
between two nodes, as is usually done in the random graph framework
(see e.g. \cite{Bollo1},\cite{Bollo2}), some authors (for certain
reasons, which come from the original idea of rewiring existing links)
made a different choice, referring the probability of a random edge
(or \tit{shortcut}) to the number of nodes, $N$, in the graph (for the
case $k=1$!). The relation between these two probabilities, $p$ and
$\phi$ is descibed at the beginning of the following section.

Important (random) graph characteristics to be employed in detecting
the small-world effect are the \tit{(expected) diameter} of the graph
and the \tit{mean-distance} between pairs of nodes. A little bit
surprisingly, it turns out to be possible to estimate or calculate
these quantities in the Strogatz-Watts-model as functions of the two
parameters, $N$ and $p$ or $\phi$. This is remarkable as one
has to deal with two coupled nonlinear difference or differential
equations for the variables $f(n)$ and $g(n)$ described in the
following section or the figure caption to figure 1. The degree of
nonlinearity varies of course depending on the extent of approximations
being made.

The general observation is that, depending on the number of shortcuts
in the network, there exist several regimes in the parameter space.
To put it more succinctly, we solve, on the one hand, the equations of
the model for fixed $p$ and $N$. On the other hand, it is interesting
to study the limit $N\to\infty$ with $p(N)$ now a function vanishing
for large $N$. That is, in this latter case, we analyse the different
asymptotic regimes around the ``point'' $(N=\infty,p=0)$. In the first
case one can study the change of behavior of the network for either
$N$ fixed and changing $p$ or vice versa.

For very small $p$ (more precisely, very few shortcuts) the average
distance, for example, scales linearly with $N$. For still quite small
$p$ one expects a (transition like) threshold or, rather, a threshold
region, above which the average distance (or the diameter) scales
roughly like in a \tit{sparse} random graph, i.e. more or less
logarithmically.  There was a certain debate about the nature of this
transition zone.  We show in the following that instead of a threshold
one actually has a relatively ``broad'' \tit{cross-over} region in
which the scaling changes in a smooth way from linear through $\sim
N^{\epsilon}[(1-\epsilon)\ln N+O(1)]$ (in first order) to $\ln N$
depending on $p$. More precisely, if we scale $p$ with $N$ and choose
$N$ large the corresponding values of the edge probability are $p\sim
N^{-2}$ (linear), $p\sim N^{-(1+\epsilon)}$ (intermediate), $p\sim
N^{-1}$ (logarithmically), respectively. The original threshold was
(in our units) conjectured to occur for $p\sim N^{-2}$.

Another interesting conjecture, which was then corroborated both
numerically and by plausibility arguments, was a
\tit{finite-size-scaling ansatz} for 
the shape of the functional dependence of the average distance, $L$,
on $N$ and $p$. We were able to confirm this ansatz modulo some
deviations which occurred in a certain region of the parameter space.

At the end of the paper we introduce a (fractal-like) notion of dimension
for networks and calculate the \tit{dimension} of the small-world-network.

To briefly characterize our own approach, we include, on the one hand,
more possibly relevant terms making hence the evolution equations more
complicated. In contrast to using then approximate solutions we manage
to derive upper and lower bound comparison difference equations
(differential equations) for the ``exact'' solutions, which allow us
to enclose them from above and below. By this method we
are able to compare the reliability of the various (approximate)
results produced in the literature, relate them to our exact bounds
and represent them in a single diagram. Last but not least we were
quite scrupulous to compensate for the possible quantitative errors
(coming from \tit{overcounting}) which are easily introduced by being
too cavalier as to the (thumb rule like) \tit{spreading argument}
frequently envoked for random graphs.

We expect that our method of providing comparison difference or
differential equations for complicated non-linear equations, which, on
their side, are presumably not solvable, may represent a strategy
which might prove useful in a more general context.  
\section{The Description of the Small World Model}
We start from the graph $\Z_N$, i.e. $N$ nodes on a line with periodic
boundary condition; that is, node $x_N$ is linked to node
$x_1$.\\[0.3cm]
Remark: To make the red thread of our analysis better visible, we
treat for the time being only the nearest neighbor model. A node $x_i$
is only connected to $x_{i\pm 1}$ (i.e. $k=1$, or coordination number
$z=2$). The more general case is a straightforward generalisation and
can be reduced to the case $k=1$ by a renormalization step, cf.
\cite{Newman1}.
\\[0.3cm]
In a next step we superimpose this graph with a true random graph,
living on the same $N$ nodes and having independent \tit{edge
  probability} $p$ (cf. for example \cite{Bollo1} or \cite{Bollo2}).
This entails that the expected number of random edges in our model is
$p\cdot N(N-1)/2$ (the average vertex degree in the respective random
graph) and the expected number of random edges being incident with a
fixed but arbitrary node, $x_i$, is $p\cdot(N-1)$.  Note that with
this definition it may happen that some of the nearest regular
neighbors of a node $x_0$, can now also be linked to $x_0$ by a random
edge. This plays however no role in the global analysis and could of
course be avoided but makes the numerical analysis more compact.

The above $p$ should be compared with the probability, $\phi$,
occurring in \cite{Newman1} or \cite{Newman2}. The latter one is
referred to the existing number of regular (non-random) edges, that is
$k\cdot N$, or $N$ for $k=1$. The reason for this derives from the
original model in which existing edges were randomly rewired. Thus,
for $k=1$, $\phi$ leads to an expected number of random edges in the
graph equal to $\phi\cdot N$ instead of $p\cdot N(N-1)/2$ in our model
($N$ large). The two probabilities are hence related by
\begin{equation}p=2\phi/(N-1) \end{equation}
if we refer them to the same global expected number of shortcuts in
the superimposed random graph.

We are in particular interested in the \tit{small world} effect. What
is usually studied is the \tit{mean distance}, $L(G)$, between two
arbitrarily selected nodes, $x_i$ and $x_j$. Note that graphs are
discrete \tit{metric spaces} in a natural way, the distance
$d(x_i,x_j)$ being given by the length of a shortest path, connecting
them (number of consecutive edges). If the individual realisations of
graphs or networks belong to a sample (probability) space, an
averaging has to be performed both over the selected pairs of nodes and the
sample space (cf. \cite{Newman1} or \cite{Requ1}).

This quantity is closely related to another important graph
characteristic, the \tit{(expected) diameter}, which we will study in
the following. Choosing an arbitrary start node, $x_0$, the graph
metric allows to define \tit{l-neighborhoods}, $U_l(x_0)$, with
\begin{equation}U_l(x_0)=\{x_i\,,\,d(x_0,x_i)\leq l\} \end{equation} 
and their respective \tit{boundaries}, defined by
\begin{equation} \Gamma_l(x_0)=\{x_i\,,\,d(x_0,x_i)= l\}    \end{equation}
With $|\Gamma_l(x_0)|$ denoting the number of nodes lying in
$\Gamma_l(x_0)$, the sequence of this values is called the
\tit{distance degree sequence} relative to node $x_0$ and is denoted
by $dds(x_0)$. When tabulating this for the full node set we arrive at
the \tit{distance distribution} $dd(G)=\{D_1,D_2,\ldots,\}$ with $D_l$
the number of pairs of nodes having distance equal to $l$
(\cite{Harary} or \cite{Requ1}). We have the following formula for the
mean distance:
\begin{equation}\label{diameter}L(G)=M^{-1}\cdot\sum_{l=1}^D l\cdot D_l     \end{equation}
with $M=N(N-1)/2$ being the number of different pairs of nodes. The
number $D=D(G)$, that is, the maximal distance occurring in this
counting is called the \tit{diameter} of the graph.

Evidently, $L(G)$ and $D(G)$ cannot be expected to be the same
numerically but in the generic situation one may surmise that they are
closely related and scale in the same way for, say, $N\to\infty$
(being motivated by the qualitative picture of \tit{spreading} in a
random graph). While the precise analytic calculation of the degree
sequence $dd(G)$, the mean distance or the diameter is a quite
ambitious task in the random graph framework (see for example
\cite{Bollo1}), the qualitative behavior can be inferred as follows.

If the edge probability, $p$, is sufficiently low, a randomly selected
node, $x_0$, has on average $p\cdot N$ neighbors and roughly $p^2N^2$
second neighbors and so on as long as the number of vertices being
reached is not to large compared to the total number $N$. If this
latter condition does no longer hold, the probability increases that
one meets a given vertex twice. Hence, due to this overcounting the
true numbers are systematically smaller, the deviations becoming
appreciable when $N$ is approached.

In the sequel we therefore employ the following strategy. Instead of
calculating the exact distance degree sequence or the exact diameter
of our small world model, we calculate, among other things, the
number of steps necessary to reach the fraction $\alpha\cdot N$ of
nodes with $\alpha$ preferably chosen as $1/2$. In this way we hope to
avoid the problems of overcounting at least to a large degree, while,
on the other hand, we expect the scaling behavior of the respective
quantities to be more or less the same as for the true numbers.
\section{The Derivation of the Evolution Equations}
As we remarked at the end of the preceding section, we want to
estimate the expected number of steps necessary to reach, for example,
half of the number of vertices, starting from a fixed but arbitrary
vertex, $x_0$. We expect that this quantity displays the same $N$- and
$p$-dependence as the mean distance or the diameter of the network or
graph under discussion, avoiding at the same time the problem of
overcounting or, on the other hand, of very complex equations when
approaching the total number of vertices, $N$.

As has been done in \cite{Newman2}, we choose the following two variables.
\begin{defi}$f(n)$ denotes the expected number of nodes, not reached
  after n steps, starting from an arbitrary but fixed node, $x_0$
  (``free nodes''). $g(n)$ denotes the number of gaps, that is, the
  number of (connected) segments of nodes, lying on the original
  $\Z_N$, not yet reached and which are separated by the segments of
  nodes already reached after $n$ steps (cf. figure 1).
\end{defi}
We note that our evolution equations describe the evolution of mean- or
expected values. In some respects this approach hence shares some
characteristics with what one calls mean-field theory in statistical
mechanics (cf. also \cite{Newman2}). However, we think, the
approximations being made by us are not so drastic as in typical
mean-field models, where, among other things, Hamiltonians are
typically strongly modified (frequently almost linearized). This is
not the case in the small-world model which, in particular in our
approach, contains strongly non-linear terms which encode at least
part of the fluctuation content in integrated form (see, for example,
the discussion about the inclusion of terms incorporating the effects
of very small gaps around eqn (10)). In a sense, what we call ``full
equations'' in the following rather describe the behavior of a
``typical'' or generic small-world graph. So it does not come as a
terrible surprise that the evolution equations for the expected values
are in relatively good agreement with what follows from real numerical
simulations of the model.

On the other hand, statistical fluctuations and correlation are not
really treated by us while this could be done in principle as the
underlying probability space is explicitly given, that is, the regular
graph $\Z_N$ superimposed with a random graph for which probability
theory is well established. One therefore may make the slightly vague
statement that the small-world model is, depending on the degree of
approximations, of an intermediate character.

To arrive at equations which are not only asymptotically correct or are
only good in a restricted region of parameter or phase space, we try
to include as many relevant terms as possible (under the proviso that
the resulting coupled and non-linear difference or differential
equations are still solvable). We start with the difference
equation, describing the expected change of the number of gaps between
consecutive steps. 

For $n=0$ we have exactly one gap, comprising all the nodes except
$x_0$, that is we have $g(0)=1$. The number of gaps increases only due
to the consecutive inclusion of shortcuts with increasing $n$,
connecting pairs of nodes in a random manner and being parametrized by
the edge probability $p$.  The main contribution in consecutive steps,
$n\to (n+1)$, comes from the term $+2pg(n)f(n)$. We will further
explain it after the introduction of equation (\ref{diffeqf}). There
exists however another contribution which acts in the opposite
direction and which becomes relevant when already many gaps do exist.
This term reads $-2g(n)^2/f(n)$ and is of a purely combinatorial (more
involved) character to be explained below when discussing equation
(\ref{diffeqf}).

The initial condition for $f$ reads $f(0)=N-1$. For $k=1$ each gap of
free nodes shrinks by two in the next step provided the gap comprises
more than one node. Neglecting in a first step this latter
possibility, the first contribution is hence of the form $-2g(n)$.
Then there is a contribution coming from new shortcuts of the form
$-2pg(n)f(n)$. The overcounting in the first term (neglection of
one-node gaps) has now to be compensated by a term $+g(n)^2/f(n)$. The
emergence of this and the corresponding term in equation
(\ref{diffeqg}) will be explained in greater detail below.
\begin{ob}\begin{align}\label{diffeqg}   g(n+1)-g(n) &= 2pg(n)f(n)-2g(n)^2/f(n)\\\label{diffeqf}    f(n+1)-f(n) &= -2g(n)-2pg(n)f(n)+g(n)^2/f(n) \end{align}
\end{ob}
We furthermore have the following apriori bounds which immediately
follow from the meaning of the respective variables in our model
system:
\begin{lemma}We have $g(n)\leq N/2$ and $g(n)\leq f(n)$.
\end{lemma}
Proof: Each gap is followed by a non-empty string of nodes being
already covered, hence the first inequality. The second one follows as
each gap contains at least one node.\hfill$\Box$\vspace{0.3cm}

The occurrence of the term $2pg(n)f(n)$ can be understood as follows.
In each step, $n\to (n+1)$, the two endpoints of each of the $g(n)$
gaps may become the source of new shortcuts to the remaining $f(n)$
\tit{free} nodes, the expected number being $pf(n)$. This leads hence
to a term of the above form in both equations. One can even be a
little bit more precise if one wants to. New gaps are not created if
the shortcuts end at free nodes which are adjacent to nodes already
reached. There are on average $2g(n)$ of them. That is, in the
equation describing the evolution of gaps the correct term is
$2pg(n)(f(n)-2g(n))$. The equation describing the evolution of $f(n)$
is not altered. This additional correction term is always negative and
we could in principle incorporate it in the following. It will make
the whole numerics slightly nastier without making a big effect. So we
will largely neglect this term but will incorporate it into what we call
the ''full equations'', see (\ref{fulldiffeqg}).

The other quadratic nonlinear terms are slightly more intricate and of
a more stochastical nature. While in equation (\ref{diffeqf}) gaps
containing only one node will contribute only one instead of two nodes
in the difference equation, in equation (\ref{diffeqg}) gaps vanish in
the next step if at level $n$ they contain only one or two nodes. The
probability for the existence of such gaps will now be calculated. We
begin with the case of one-gaps. We will solve this problem with the
help of the well-known partitioning problem of a given set into
disjoint subsets.  We associate the set of $g$ gaps and $f$ free nodes
with $f$ balls to be distributed over $g$ boxes. In general there
exist
\begin{equation}\binom{f+(g-1)}{g-1}=\binom{f+(g-1)}{f}\end{equation}
combinations (see \cite{Renyi} or \cite{Feller}). In effect we
calculate the number of different words of length $f+(g+1)$ consisting
of $(g+1)$ bars and $f$ dots, under the proviso that each word begins
and ends with a bar and that each consecutive pair of bars is divided
by a non-empty string of dots as in our case no box can be empty. This
implies that we can place exactly one ball in each box and perform the
above calculation, (7), which represents the number of partions
without the constraint of non-empty boxes, for the remaining number of
$(f-g)$ balls, yielding
\begin{equation}A:=\binom{f-1}{g-1}     \end{equation}
configurations. This is the cardinality of the set of elementary
events in our probability space.

To calculate the expected number of gaps containing only one or two
nodes, we introduce the following random variables, $Y_1,Y_2$ over the
probability space of words, we associated with the random graphs $G_j$:
\begin{equation}Y_i(G_j):=\#\{\textnormal{gaps of length }i\}     \end{equation} 
in each of the above $A$ configurations (graphs), $G_j$. Before we
proceed a short remark as to the probabilities of the individual
configurations should be in order.

In our model probability space a regular graph is superimposed by a
random graph with edge probability $p$. In our above calculation we
deal with fixed numbers $f$ and $g$. As the gaps arise due to the
existence of random edges, the gaps are expected to be randomly
scattered over the regular graph $\Z_N$ in basically the same way as
pairs of nodes are linked by random edges, that is, almost
independently. This then should also essentially hold for the number
of gaps, met after $n$ steps. Furthermore, this reasoning should not
be affected in a serious way by the possible annihilation of gaps for
large step-number $n$, as long as we stay away from the regime where
the \tit{spreading argument} for random graphs is no longer correct.
From this we see that it is a reasonable strategy to remain below the value
$N/2$ with the step number $n$. We hence conclude that in the
indicated regime each configuration should have the independent
probability $A^{-1}$.

We therefore have
\begin{equation}E(Y_i)=A^{-1}\cdot\sum_j Y_i(G_j)    \end{equation}
For the one-gaps we can represent $Y_1$ by more elementary random
variables, $y_k$ with $k$ running from $1$ to $g$, enumerating the
existing $g$ gaps and $y_k=1$ if gap $(k)$ contains only one element
and zero else. This yields
\begin{equation}Y_1(G_j)=\sum_k y_k(G_j)    \end{equation}
and
\begin{equation} E(Y_1)=\sum_k E(y_k)= g\cdot\binom{f-2}{g-2}/\binom{f-1}{g-1}    \end{equation}
where $\binom{f-2}{g-2}$ is the number of configurations with only one
element in gap $(k)$.
Correspondingly we get for the expected number of $2$-gaps:
\begin{equation}E(Y_2)=g\cdot\binom{f-3}{g-2}/\binom{f-1}{g-1}    \end{equation}
For the one-gaps this yields
\begin{equation}E(Y_1)=g\cdot g/f    \end{equation}
For the two-gaps we have
\begin{equation}E(Y_2)=g\cdot\frac{(g-1)\cdot(f-g)}{(f-1)\cdot(f-2)}    \end{equation}
and for $f$ large, i.e. $f\approx (f-1)\approx (f-2)$:
\begin{equation}\label{E}E(Y_2)\approx g\cdot( g/f-1/f-g^2/f^2 +g/f^2)    \end{equation}
With our apriori estimate $g\leq f$ and as we start from the initial
conditions $f(0)=(N-1)\approx N\,,\,g(0)=1$
in most of our phase space the dominant contribution comes from the
term $g\cdot g/f$ also in the case of two-gaps. 
\begin{conclusion} The expected number of one- or two-gaps is
  approximately
\begin{equation}E(\#(1\text{- or } 2\text{-gaps}))\approx 2g^2/f   \end{equation}
\end{conclusion}
(Note that in our probability space the possibility of being a 1-gap
or a 2-gap is mutually exclusive). This result explains the occurrence
of the correction terms in our evolution equations.  Without these
approximations, the equations (\ref{diffeqg}) and (\ref{diffeqf}) would
read
\begin{align}\label{fulldiffeqg} g(n+1)-g(n) &= 2pg(f-2g)-\frac{g\cdot
    (g-1)}{f-1}-\frac{g\cdot(g-1)\cdot(f-g)}{(f-1)\cdot(f-2)}\\
\label{fulldiffeqf} f(n+1)-f(n) &= -2g-2pgf+\frac{g\cdot
  (g-1)}{f-1} \end{align} with $f$ and $g$ instead of $f(n)$ and
$g(n)$ on the rhs. We'll refer to these as ``full equations'' and
compare them with the simplified ones ((\ref{diffeqg}) and
(\ref{diffeqf})) in figure \ref{diagfexamples}.  Note in particular
that we approximated the last term occurring in \ref{fulldiffeqg} by
$-g^2/f$ in \ref{diffeqg}, neglecting the positive higher order
contribution, being essentially of the form $+g^3/f^2$.

A brief comment is in order as to the corresponding formulas derived
in \cite{Newman2} (cf. their formulas (3),(6) or (A10),(A11)). We
decided to neglect all terms in (\ref{E}) except the leading one,
$g^2/f$, which is reasonable in our view. In \cite{Newman2}, in the
corresponding equation an additional term of the type $g/f$ occurs
(derived by a different argument). On the other hand, more important
in our view is equation (\ref{diffeqf}), which comprises three terms in
our approach (including a nasty non-linear one, $g^2/f$), while in
\cite{Newman2} only the first one, $-2g$, occurs on the rhs. This
makes the corresponding equations of course much easier to solve but
may only be a good approximation in a restricted regime of parameter
space (see the brief discussion at the end of section
\ref{quantitative}). We discuss and compare the numerical results in
section \ref{scaling}. One can see that the solution of \cite{Newman2}
is similar to our \tit{lower bound}-equation for $f$ (eqn
(\ref{ftildeeqn})), which is reasonable as in our lower bound for $f$
the quadratic term is largely suppressed (cf. the following section
and figure \ref{diagscalingfunction}).  
\section{Solution Strategies}
The above system of evolution equations contains non-linearly coupled
quantities and can be solved only in very exceptional lucky
circumstances. Instead of making more or less uncontrollable
approximations, we develop the following strategy. We try to enclose
the above exact equations by \tit{comparison equations} bounding the
exact solutions, $f(n)\,,\,g(n)$, from above and below, the
corresponding variables being denoted by 
\begin{equation}\fo (n)\,,\,\fu (n)\,,\,\go (n)\,\,\gu (n)    \end{equation}

The problem is that, on the one hand, the comparison equations have to
be so chosen that they can be rigorously solved and, on the other
hand, these bounds have to be quite good so that we are able to infer
something relevant also for the enclosed exact equations in particular
for the scaling limit $N$ very large and $p$ a vanishing function of
$N$. A central role will be played by the value $n_*$ for which we have
reached on average $N/2$ of the nodes when starting from an arbitrary
but fixed node $x_0$ (or more generally $\alpha N$ nodes with
$0<\alpha<1$). In other words, the range of $n$-values we are using is
restricted by
\begin{equation}n\in [0,n_*]\quad \text{so that}\quad (N-1)=f(0)\geq f(n)\geq
\alpha N\end{equation}

We now study the rhs of the equations (\ref{diffeqg}), (\ref{diffeqf}).
For one, we assume that these equations have been solved for our
initial conditions $g(0)=1\,,\,f(0)=N-1$, so that $g(n)\,,\,f(n)$ now
represent particular functions of $n$. On the other hand we can regard
the rhs (dropping the dependence on the variable $n$) as functions on
the phase space, spanned by the possible values of the variables
$g\,,\,f$. In our assumed range of possible parameters and variables
we have the estimate
\begin{equation}2pgN-2g^2/N\geq 2pgf-2g^2/f\geq 2pg\cdot(\alpha
  N)-2g^2/(\alpha N)             \end{equation}

The idea is now to introduce the comparison difference equations
\begin{equation}\go (n+1)-\go (n):= 2pN\go (n)-2\go (n)^2/N         \end{equation}
and 
\begin{equation}\gu (n+1)-\gu (n):= 2p(\alpha N)\gu (n)-2\gu
  (n)^2/(\alpha N)            \end{equation}
with the initial conditions
\begin{equation}g(0)=\go (0)=\gu (0)=1        \end{equation}
For the initial differences we have
\begin{equation}\go (1)-\go (0)>g(1)-g(0)>\gu (1)-\gu (0)     \end{equation}

Our strategy is now to use these comparison equations to learn
something about the true equations. Unfortunately matters are not so
transparent for difference equations as compared to differential
equations. The reason is that they are only given at discrete points
and may (therefore) display a more complex behavior (see for example
Hoelder`s theorem and extensions thereof in \cite{Noerlund},p.283 or
\cite{Meschkowski},p.220). Due to these problems we will in the
following go over to the corresponding differential equations, being
however aware of the fact that it does not seem to be an easy task to
provide good error estimates, in particular as the differences in our
context are not infinitesimal (as to this interesting question of
principle cf. the discussion in \cite{Bender}). What is furthermore
remarkable is the observation (see below) that we can prove a useful
theorem in the case of differential equations the analogue of which,
as far as we can see, we cannot prove for difference equations (at
least with the same methods).

The corresponding differential equations read:
\begin{align}\label{differentialg}g'(x) &= 2pg(x)f(x)-2g(x)^2/f(x)\\
\label{differentialf}  f'(x) &= -2g(x)-2pg(x)f(x)+g(x)^2/f(x)
\end{align}
with $g(0)=1\,,\,f(0)=(N-1)\,,\,x\in [0,x_*]$ so that $(N-1)\geq
f(x)\geq \alpha N$. The comparison differential equations with respect
to $g(x)$ are
\begin{align}\go'(x) &= 2pN\go (x)-2\go(x)^2/N\\
\gu'(x) &= 2p(\alpha N)\gu (x)-2\gu(x)^2/(\alpha N)
\end{align}
with $\go(0)=\gu(0)=1$.

As to equation (\ref{differentialf}) we proceed as follows (we note that
we in fact experimented with different possibilities; the one we are
presenting below seems to be the most appropriate one). We mentioned
above the apriori bound $g\leq f$. For the rhs of
equation (\ref{differentialf}) we then have:
\begin{equation}-2g-2pgf+g^2/f\leq -2g-2pgf+g=-g-2pgf
\end{equation}
On the other hand we have also
\begin{equation}-2g-2pgf+g^2/f>-2g-2pgf      \end{equation}
Therefore our comparison differential equations for $f$ are
\begin{align}\fo'(x) &= -\gu(x)-2p\gu(x)\fo(x)\\
  \fu'(x) &= -2\go(x)-2p\go(x)\fu(x)\end{align}
with $\fo(0)=N\,,\,\fu(0)=N-2$.\\[0.3cm]
Remark: As $N$ is supposed to be very large, it does not make a big
difference for the numerical calculations to let all the initial
values be equal to $N$. The above choice makes the analytic
argument a little bit simpler (see below).\vspace{0.3cm}

To further exploit these comparison equations we proceed as follows.
Note that we have succeeded in decoupling the equations for $f$ and
$g$. We can solve the differential equations for $\go$ and $\gu$ and
plug the solutions into the $\fo$- and $\fu$-equation. We assume that
together with the comparison differential equations the original
differential equations for $g$ and $f$ have been solved for the
mentioned initial conditions. In the differential equation for $g$
with initial condition $g(0)=1$ we can then regard the corresponding
$f$-solution as an external function with $g(x)$ solving this ``new''
differential equation (together with the given initial condition). We
compare the solution $g(x)$ of this latter equation with the solutions
of the differential equations for $\go\,,\,\gu$ respectively. We have
\begin{equation}\go(0)=\gu(0)=g(0)=1\quad\text{and}\quad \gu'(0)<g'(0)<\go'(0)     \end{equation}

We now prove the following result.
\begin{prop}Let $y_1(x)$ and $y_2(x)$ be solutions of the two
  differential equations
\begin{equation}y_1'(x)=F_1(y_1,x)\;,\;y_2'(x)=F_2(y_2,x)    \end{equation}
on the interval $[0,x_*]$ with $y_1(0)\leq y_2(0)$. Let $F_1\,,\,F_2$
fulfil
\begin{equation}F_1(y,x)<F_2(y,x)    \end{equation}
on the domain $[0,x_*]\times I_y$, $I_y$ a suitable $y$-interval and
with both $y_1(x)\,,\,y_2(x)$ staying in this domain. Then
\begin{equation}y_1(x)\leq y_2(x)\quad\text{on}\quad [0,x_*]     \end{equation}
\end{prop}
Proof: From the assumptions it follows that $y_1(x)< y_2(x)$ in some
open interval $(0,\varepsilon)$. If $y_1(x)> y_2(x)$ for some $x$,
there exists an $r>0$ (by continuity) with $y_1(r)= y_2(r)$ and
$y_1(x)> y_2(x)$ in an open interval $(r,r+\varepsilon')$. But this is a
contradiction since
\begin{equation} F_1(y_1(r),r)<F_2(y_1(r),r)    \end{equation}
hence again implying that $y_1(x)<y_2(x)$ in an open interval
$(r,r+\varepsilon'')$. We conclude that $y_1(x)\leq y_2(x)$ on
$[0,x_*]$.\hfill$\Box$

It sometimes happens that we have $F_1(y,x)<F_2(y,x)$ on the open
interval $(0,x_*)$ but $F_1(y(0),0)=F_2(y(0),0)$ for some value
$y(0)$. We then can prove the following corollary:
\begin{koro}Making the same assumptions as before except for
  $F_1(y(0),0)=F_2(y(0),0)$ instead of $F_1(y(0),0)<F_2(y(0),0)$. We assume that there exists a parameter
  $\lambda$ so that on the closed interval we have
\begin{equation}F_1(y,x;\lambda)<F_2(y,x)    \end{equation}
for $\lambda>0$ and $F_1(y,x;0)=F_1(y,x)$, the dependence on $\lambda$
being continuous or differentiable. Then the parameter dependent
solutions converge pointwise towards the solutions for $\lambda=0$
(cf. \cite{Coddington}). For $\lambda>0$ our above result applies. So,
by continuity it applies also in the limit $\lambda\to 0$.
\end{koro}
Note that in our case the parameter $\lambda$ is the parameter taking
the values $N,N-1$ etc.\\[0.3cm]
Remark: We surmise that such comparison results are known in the large
literature about differential equations but we were unable to find a reference.
\begin{conclusion} What we have now shown is that under the
  assumptions being made, $\go\,,\,\gu$ and $\fo\,,\,\fu$ are upper and
  lower bounds of the corresponding solutions $g\,,\,f$ of the
  original differential equations. 
\end{conclusion}
\section{\label{quantitative}The Quantitative Results}
With our $\alpha$ being now either 1 or $\frac{1}{2}$ we can express both the
upper and lower bound by a single equation:
\begin{eqnarray}
\tilde{g}'(x) &=& 2\,\alpha\,p\,N\,\tilde{g}(x) -
\frac{2\,\tilde{g}(x)^2}{\alpha\, N}
\end{eqnarray}
with $\tilde{g} = \overline{g}$ for $\alpha = 1$ and $\tilde{g} =
\underline{g}$ for $\alpha = \frac{1}{2}$. This nonlinear differential
equation (of Bernoulli type) can be transformed into a linear one with
the help of the transformation $z := \tilde{g}^{-1}$ and yields with
the initial condition $\tilde{g}(0) = 1$ the solution
\begin{eqnarray}
\tilde{g}(x) &=& \left(
  \left(1-\frac{1}{p\,\alpha^2\,N^2}\right)\,e^{-2\,p\,\alpha\,N\,x} +
  \frac{1}{p\,\alpha^2\,N^2}\right)^{-1}
\end{eqnarray}
Inserting these solutions into the corresponding upper and lower bound
equations for $f(x)$
\begin{eqnarray}
\overline{f}'(x) &=& - 2\,p\,\underline{g}(x)\,\overline{f}(x) -
\underline{g}(x)\\
\underline{f}'(x) &=& - 2\, p\, \overline{g}(x)\, \underline{f}(x) -
2\,\overline{g}(x)
\end{eqnarray}
and introducing the parameter $\beta \in \{1,2\}$ so that $\tilde{f}'
= -2\,p\,\tilde{g}\,\tilde{f} - \beta \tilde{g}$ we obtain (after a
simple variable transformation, $\overline{f}\rightarrow \overline{f} +
1/2p$, $\underline{f}\rightarrow \underline{f} + 1/p$, and separation
of variables) the following result, using $f(0) = N$ as initial
condition instead of $N-1$:
\begin{eqnarray}\label{ftildeeqn}
\tilde{f}(x) &=& \frac{2\,p\,N + \beta}{2\,p}\left(1 - \frac{1}{p\,\alpha^2N^2}
+ \frac{1}{p\,\alpha^2N^2}\,e^{2\,p\,\alpha\,N\,x}\right)^{-p\,\alpha\,N} -
\frac{\beta}{2\,p}
\end{eqnarray}
We are interested in the value $\tilde{x}_*$ of $x$ for which $\tilde{f} =
\frac{1}{2}N$. Solving for $\tilde{x}_*$ yields
\begin{eqnarray}
\tilde{x}_* &=& \frac{1}{2\,\alpha\,p\,N}\,\ln\left(\left[\left(\frac{2\,p\,N +
        \beta}{p\,N + \beta}\right)^\frac{1}{p\,\alpha\,N} -
        1\right]\,p\,\alpha^2N^2 + 1\right)
\end{eqnarray}
the lower bound being assumed for $\alpha = 1, \beta = 2$ the upper
bound for
$\alpha = \frac{1}{2}$ and $\beta = 1$.\\[0.3cm]
Remark: The argument of the logarithm is always larger than one, as
$((2pN+\beta)/(pN + \beta))^{1/p\alpha N}>1$ and $p\alpha^2 N^2 > 0$.
We furthermore want to stress the important point that a too poor
approximation of, say, the upper bound, $\overline{f}(n)$, may easily lead
to a function which does not decay sufficiently. In that case our
estimates would have been useless. We see however that our above
choice is strong enough. \vspace{0.3cm}

We now want to investigate the scaling regime $N\rightarrow \infty$,
$p = c N^{-1-\epsilon}$ with $\epsilon \in [0,1], c > 0$, with the
boundary cases $p\propto N^{-1}$ and $p\propto N^{-2}$ being
particularly interesting.
Inserting these choices into the preceding equation we get
\begin{eqnarray}
\tilde{x}_* &=& \frac{N^\epsilon}{2\,\alpha\,c}\,\ln\left(\left[\left(\frac{2\,c\,N^{-\epsilon} +
        \beta}{c\,N^{-\epsilon} + \beta}\right)^\frac{N^\epsilon}{c\,\alpha} -
        1\right]\,c\,\alpha^2N^{1-\epsilon} + 1\right)
\end{eqnarray}
For $\epsilon = 0$ we have
\begin{eqnarray}
\tilde{x}_* &=& \frac{1}{2\,\alpha\,c}\ln N + C_1(\alpha,\beta,c)
\end{eqnarray}
with $C_1$ being of the precise form:
\begin{equation}C_1=(2\alpha\,c)^{-1}\cdot\ln\left(c\alpha^2\left[\left(\frac{2c+\beta}{c+\beta}\right)^{1/c\alpha}-1\right]+1/N\right)
\end{equation}
which becomes asymptotically independent of $N$ for large $N$.

For $\epsilon \neq 0$ we have (developing
the logarithm up to the first order)
\begin{eqnarray}
    \left(\frac{2\,c\,N^{-\epsilon} + \beta}{c\,N^{-\epsilon} +
    \beta}\right)^\frac{N^\epsilon}{c\,\alpha} 
&=& \exp\frac{N^\epsilon}{c\,\alpha}\,
    \left[\ln\left(\frac{2\,c}{\beta\,N^\epsilon} + 1\right) -
    \ln\left(\frac{c}{\beta\,N^\epsilon} + 1\right)\right]\\
&=& \exp\frac{N^\epsilon}{c\,\alpha}\left[\frac{c}{\beta\,N^\epsilon} +
    O(N^{-2\epsilon})\right]\\
&=& \exp\left(\frac{1}{\beta\,\alpha} + O(N^{-\epsilon})\right) 
\end{eqnarray}
 and get
\begin{eqnarray}
\tilde{x}_* &=& \frac{N^\epsilon\,\left(\left(1-\epsilon\right)\ln N +
    \ln\left(\left[\exp(1/\alpha\beta)-1\right]c\alpha^2+1/N^{1-\epsilon}
    \right) \right)}{2\alpha c}
\end{eqnarray}
which obviously also describes the boundary cases $\epsilon = 0$ and
$\epsilon = 1$ (provided we would include the neglected term
$O(N^{-\epsilon})$ which is now $O(1)$). This behavior is valid both
for the upper and lower bound of $f$. As the $x_*$ for the true $f$
has to lie between the respective values for the upper and lower bound
we infer that it has the same scaling behavior for
$N\rightarrow\infty$.
\begin{conclusion} We infer that between $p\varpropto N^{-1}$ and $p\varpropto
  N^{-2}$ there exists a broad transition zone with the scaling of the
  diameter or mean distance exactly interpolating between these two boundary
  cases, $x_* \approx \ln N$ and $x_* \approx N$ (up to now we have
  only studied the scaling of $x_*$, a parameter which is of course
  closely related to the above mentioned graph characteristics; see below).
\end{conclusion}

In our above calculations we dealt with the expected value, $x_*$, at
which the expected number of free nodes drops to the value $N/2$. We
argued above that the corresponding (exact) formulas for the value, at
which this number assumes the value zero, would be much more
complicated. To make nevertheless a statement about this value we
apply the following (plausibility) argument (which, however, should
not be viewed as a rigorous proof). Put differently, we will provide
an argument which is expected to hold only for \tit{expectation
  values} or \tit{typical} nodes.  Let $X$ be an arbitrary initial
vertex and $X'$ a vertex with largest possible distance to $X$ in a
given realisation of the network. The expected radius of the
$N/2$-neighborhoods for both vertices is equal to $x_*$. In case the
corresponding $x_*$-neighborhoods $U_*(X)$ and $U_*(X')$ of $X$ and
$X'$ are not disjoint the distance between $X$ and $X'$ must be less
than $2\,x_*$.  If these neighborhoods are disjoint the graph $G$ is a
disjoint union of $U_*(X)$ and $U_*(X')$ and hence the distance
between $X$ and $X'$ must be exactly $2\,x_*$.  As all these arguments
apply only to the generic case, we can associate this value with the
\tit{expected diameter} of our network, denoted by $D$ and get the
estimate
\begin{equation}\label{D}x_*\leq D\leq 2x_*     \end{equation}
Remark: We again emphasize that this argument is only correct in an
averaged sense in which all nodes are assumed to stand on the same
footing. It is of course easy to design particular graphs where this
estimate does not hold. Take for example a graph having a densely
entangled neigborhood around some node $x$ from which a long
one-dimensional string emanates. In this case the diameter is of
course much larger than $2x_*$. We think one could prove something
rigorous at this place, which, on the other hand, may be a little bit
tedious and unnecessarily blow up the paper.\vspace{0.2cm}

On the other hand we infer from equation (\ref{diameter}) that the
average distance, $L$ fulfils
\begin{equation}L\leq D    \end{equation}
Taking again a typical node, $X$, (so that its neighborhood $U_*$
has approximately $N/2$ members), we can approximate the mean distance
$L$ by $(N-1)^{-1}\cdot\sum l\cdot |\Gamma_l|$ and get the estimate
\begin{equation}L\geq (N-1)^{-1}\cdot\left(\sum_1^{x_*-1}l\cdot
    |\Gamma_l|+N/2\cdot x_*\right)\geq x_*/2  \end{equation}
as 
\begin{equation}\sum_{l\geq x_*}l\cdot |\Gamma_l|\geq
  x_*\cdot\sum_{l\geq x_*}\cdot |\Gamma_l|\geq x_*\cdot N/2\end{equation} 

We thus get
\begin{equation}x_*/2\leq L\leq D\leq 2\,x_*    \end{equation}
We can now insert the respective parameters, $\alpha$ and $\beta$ in
our expressions for $\tilde{x}_*$, thus yielding
\begin{equation}x_*^{\alpha = 1,\beta = 2}/2 \leq x_*/2 \leq L \leq
  D\leq 2\,x_*  \leq
  2\,x_*^{\alpha = \frac{1}{2}, \beta = 1} \end{equation}
With  our numerical expressions for the upper and lower bounds
we get
{\small
\begin{equation}  \!\!\!
    \frac{1}{4pN}\ln{
    \left(\!\left[\!\left(\frac{2pN\!+2}{pN\!+2}\right)^{\!\!\frac{1}{pN}}
    \!\!\!\! - 1\right]pN^2 + 1\!\!\right)} \leq L \leq
    \frac{2}{pN}\ln{
    \left(\!\left[\!\left(\frac{2pN\!+1}{pN\!+1}\right)^{\!\!\frac{2}{pN}}
    \!\!\!\! - 1\right]\frac{pN^2}{4} + 1\!\!\right)}
\end{equation}}

For very small ($\epsilon >1$ such that $pN^2\to 0$) or vanishing $p$
the inequality reduces to $0.16\,N \leq L \leq 3.19\,N$, with
$L=0.25\,N$ for the true $L$ in the case $p=0$. For $p =
cN^{-1-\epsilon}$, $c$ of order one and $\epsilon \in [0,1]$, and a
reasonable number of shortcuts ($pN^2 > 1$), $L$ displays a behavior already
exemplified for $x_*$:
\begin{equation}
L \approx \frac{C_2\, N^\epsilon}{c} \left(\left(1-\epsilon\right)\ln N + \ln
  c + C_1(c)\right)
\end{equation}
with $C_1$ constant for $\epsilon \neq 0$. For very strongly connected
graphs ($\epsilon \in [-1,0)$), these bounds become invalid, as $L$
tends to one and the differential equations will no longer approximate
the difference equations well enough.

We want to come back to the question of the importance of the
non-linear quadratic terms in our evolution equations (\ref{diffeqg})
and (\ref{diffeqf}). One may be led to the wrong conclusion that they are
always marginal because initially $g$ is very small compared to the
huge $f$. But one should note that $g$ grows very fast for certain
choices of the parameter $p$. To get some feeling we take, for
example, $\overline{g}$ and ask for what values of $x$
$\overline{g}^2/N$ is of order $\overline{g}$.

This is the case if $\overline{g}\approx N$. The result strongly
depends on the value of $p$. Inserting $p=c/N$ in the equation for 
$\overline{g}$ and solving for $x$ we get $x\approx (2c)^{-1}\cdot\ln
N$. Hence the quadratic contribution becomes appreciable when $x$
approaches the regime where $f$ drops to $N/2$, that is, the regime we
are interested in. Put differently, it is dangerous to neglect this
term on apriori grounds.

On the other hand, taking for example $p=1/N^2$, and making the same
calculation we infer that the non-linear term remains negligible in
the domain we are interested in. For $\epsilon>0$ we get of course
intermediate results.

The effects of neglecting the even smaller combinatorial terms which
appear in the full eqns (\ref{fulldiffeqg}) and (\ref{fulldiffeqf}) can be
seen in figure \ref{diagfexamples}: Here the neglection has a greater
effect for the case $p=1/N^2$ than for $p=1/N$, as the total number
$g$ of gaps in the first case is smaller and therefore nearer to 1
than in the second case. Nevertheless,these discrepancies are still
negligible.


\section{\label{scaling}Comparison with former results}

Barth\'el\'emy and Amaral, and later also Newman and Watts
(\cite{Amaral1},\cite{Newman1}) conjectured a scaling behavior for $L$
of the form
\begin{equation}
L = N\cdot F(pN^2)
\end{equation}
with some universal function $F$ with $F(y) \rightarrow \frac{1}{4}$
for $y\rightarrow 0$ and $F(y) \rightarrow C\,\ln(y)/y$ for
$y\rightarrow \infty$. Our bounds do not scale exactly in this way,
but at least do so approximately for large $N$ and $\epsilon > 0$. In
this regime the $pN$-dependend term in the logarithm tends to the
constant $\exp(1/\alpha\beta)$ and our bounds assume the form
\begin{equation}
  L = N\cdot F(pN^2) \quad \textnormal{with} \quad
  \frac{1}{4\,y}\,\ln{\left(\left[e^{1/2} - 1\right]y + 1\right)} 
  \leq F(y) \leq
  \frac{2}{y}\,\ln{\left(\frac{e^2 - 1}{4}\,y + 1\!\!\right)}   
\end{equation}
On the other hand, for large $pN$ and $\epsilon < 0$ or $\epsilon = 0$
with large $c$, this scaling behavior breaks down and our
estimate for the average distance scales like $(\ln N)/pN$. This
estimate makes however only sense for $\ln N > pN$, as $L\geq 1$. Thus
the case $\epsilon < 0$ isn't described correctly by this formula. For
$\epsilon = 0$, $L$, according to this scaling-ansatz, simply scales
like $(\ln N)/c$, without any correction term of the form $(\ln c)/c$,
which occurs in our above presumably more exact result. So, although
$L$ correctly scales with $\ln N$ in both cases (depicting a random
graph), there exist certain deviations for large $c$.

In \cite{Newman2}, Newman, Moore and Watts found the following
expression for their universal scaling function
\begin{equation}
F_{NMW}(y) = \frac{1}{2\sqrt{y^2+2y}}\tanh^{-1}\sqrt{\frac{y}{y+2}}.
\end{equation}
For $p\propto N^{-2}$ ($\epsilon = 1$) this function is a constant. On
the other hand, for $p = cN^{-1-\epsilon}$ with $0 \leq \epsilon < 1$,
the argument $y=pN^2$ is large for large $N$ and
\begin{equation}
  L = N\,F_{NMW}(pN^2) \approx \frac{\ln p\,N^2 + \ln 2}{4\,p\,N}
  \approx \frac{N^\epsilon((1-\epsilon)\ln N + \ln c + \ln 2)}{4\,c}
\end{equation}
In figure 3 $F_{NMW}$ and the scaling functions
deriving from our bounds are shown, indicating that the NMW-ansatz
complies with them. Note however that this is only valid for $\epsilon > 0$ or
$\epsilon = 0$ with a not too large $c$ in $p=c/p$. 

In \cite{Barbour}, Barbour and Reinert made a rigorous analysis of the
probability distribution for the distance function, getting the
following result: Let $X$ and $X'$ be some randomly chosen vertices on $G$,
$\rho = pN$ and $S := pN^2$
(to be identified with $L\rho$ in \cite{Barbour}) then
\begin{equation}
P\left[d(X,X') > \frac{\ln S}{2\,\rho} + \frac{x}{\rho}\right] =
\int_0^\infty\frac{e^{-y}\,\textnormal{d}y}{1+y\,e^{2x}} +
O\left(\frac{e^x(1+e^{2x})\ln^2 S}{\sqrt{S}}  \right)
\end{equation}
for all $-\frac{1}{2}\ln S \leq x < \frac{1}{4}\ln S$. $L$ is the mean
distance resulting from this distribution for $d(X,X')$. To make
things simpler, we instead treat the median of it, which can be easily
approximated by the special choice $x = 0$. Neglecting the error term
we obtain
\begin{equation}
P\left[d(X,X') > \frac{\ln S}{2\,\rho}\right] \approx 0,596
\end{equation}
stating $L \approx L_\textnormal{median} \approx \ln(pN^2) / 2pN$ or,
with $p = cN^{-1-\epsilon}$,
\begin{equation}
L \approx \frac{N^\epsilon\,((1-\epsilon)\ln N + \ln c)}{2c}.
\end{equation}
The corresponding universal scaling function reads $F_{BR} = (\ln y)/2y$,
and is also depicted in figure 3. For $pN^2 > 2$ (at least one
expected shortcut) this lies within our bounds. Below this, the error term of
the probability $P$ rises above one and $F_{BR}$ looses its meaning.

\section{Dimension of the Small World}

In \cite{NowRequ}, two related dimensional concepts (of a fractal
type) were introduced for infinite graphs (note the close connection
to the \tit{distance degree sequence}, discussed in \cite{Requ1}) and
a number of its properties proved. We learned later that this concept
occurred already earlier in the literature but, as far as we can see,
its interesting properties were never systematically studied (see for
example \cite{Baxter}). A technically different but physically related
concept was exploited by Dhar (\cite{Dhar}), see also \cite{Filk}.
Furthermore Ising models on such irregular spaces were studied
recently, an early source being \cite{Scalettar}.  The reason to deal
with infinite graphs is that only in the limit $N\to\infty$ the global
notion of dimension becomes independent of local (model dependent)
aspects like e.g. coordination numbers of, say, lattices, all having
the same embedding dimension. One of the two definitions reads:
\begin{defi}
  Let $G$ be an arbitrary graph with $N$ vertices and $U_l(x)$ the
  $l$-neighborhood of the vertex $x\in G$. Then we define the {\em
    dimension} of $G$ (relative to $x$) as
\begin{equation}
\dime_x(G) := \lim_{l\rightarrow \infty} \frac{\ln \,\#U_l(x)}{\ln l}
\end{equation}
(provided the limit exists; in general we have to deal with $\liminf$
and $\limsup$).
\end{defi}
In \cite{NowRequ} it was shown, that this notion of dimension (also
called the ``internal scaling dimension'') is independent of the
initial vertex $x$ (under a mild technical assumption). For finite
(but large) and connected graphs we can instead employ the following graph
characteristic:
\begin{equation} \dime_{approx}(G) = \ln N / \ln \diam(G)\end{equation}
For real networks this value has been analysed in
\cite{CsanyiSzendroi}. There one can see, that the approximate
dimension might tend to underestimate other notions of dimension,
because $\ln \#U_l(x)$ saturates for large $x$ (an effect which is
however pretty obvious as the spreading argument does of course no
longer hold in that regime). On the other hand, in our case we expect
$\ln \#U_l(x)$ not to saturate before reaching $N/2 =: \#U_l(x_*)$
(cf.  figure \ref{diagfexamples}). As $x_*\leq \diam G \leq 2x_*$ (see
however the discussion after eqn(\ref{D})), we
get
\begin{eqnarray}
\frac{\ln N - \ln 2}{\ln x_* + \ln 2}\leq \dime_{approx} (G) \leq \frac{\ln
  N}{\ln x_*}.
\end{eqnarray}
Thus, for large enough $N$ and $x_*$ the approximate dimension doesn't
deviate
too much when staying below $N/2$ and won't suffer from
saturating-effects.

Applying this concept to the Small World Model for large $N$ we get
\begin{equation}
\dime_{approx}(G) = \frac{\ln N}{\ln \diam(G)} = \frac{\ln N}{\ln
  C_1\,N^\epsilon (C_2 + \ln N^{1-\epsilon})}
\end{equation}
which is $\approx \frac{1}{\epsilon}$ for $\epsilon>0$,
the constants $C_1$ and $C_2$ being independent of $N$ (depending only on
$c, \alpha$ and $\beta$). For $\epsilon = 1$ ($p\sim
N^{-2}$) we get the value one, which is reasonable for this rarefied
linear case.
For the opposite case, $\epsilon = 0$ ($p\sim N^{-1}$), we have
\begin{equation}\dim_{approx}(G)\approx \ln N/\ln\ln N       \end{equation}
which diverges for $N\to\infty$.

In \cite{Newman1}, Newman et.al. introduced a renormalization process
for the Small World Model. This process divides the graph into
segments of length 2 and interprets these segments again as vertices
in a new Small World Model with size $N' = N/2$ and
edge-probability\footnote{Note that our notation differs from that in
  \cite{Newman1}.} $p' = 4p$. With $p = c\,N^{-1-\epsilon}$ this
substitution yields $p' = c'\,N'^{-1-\epsilon}$ with $c' =
2^{1-\epsilon}c$. Hence our $\dime_{approx}=1/\epsilon$ is constant
under this renormalization. A similar phenomenon was observed in the
renormalization process for infinite graphs (with globally bounded
node degree) introduced in \cite{RequRen}.

\section{Conclusion}

We found the mean distance $L$ of a Small World Model with $N\gg 1$
nodes and edge-probability $p=cN^{-1-\epsilon}$ to be bounded from
above and below by two expressions of the form $C_2 N^\epsilon (\ln
N^{1-\epsilon} + C_1)$ (the constants depending on whether the upper
or lower bound is taken). This implies a broad transition zone, in
which the mean distance drops from a linear growth to a logarithmic
one, permitting each power law $L\sim N^\epsilon$ for $\epsilon\in
(0,1)$. Furthermore, $1/\epsilon$ can be regarded as an approximative
dimension of the corresponding graph. Our results partly corroborate
earlier work but lead also to certain numerical deviations.\\[1cm]
{\small Acknowledgement: We thank the unknown referees for the many
  valuable comments which helped to improve the paper.}

\newpage
\centerline{\large{\bf Figures}}
\begin{figure}[h]
\centerline{\epsfig{file=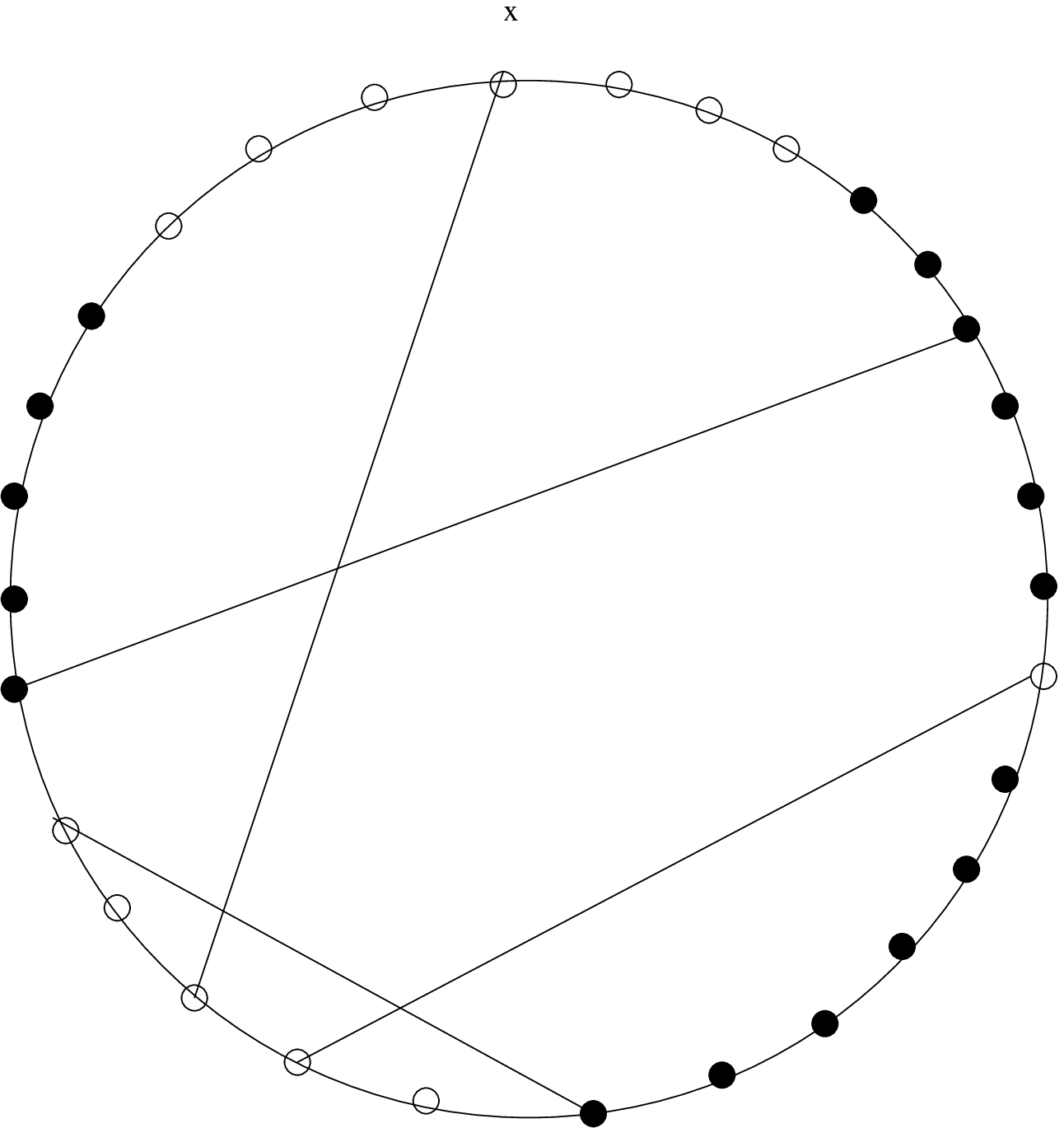,width=6cm,height=6cm,angle=0}}
\caption{}
\label{diagsmallworld}
\end{figure}

\begin{figure}[!ht]
\centerline{\epsfig{file=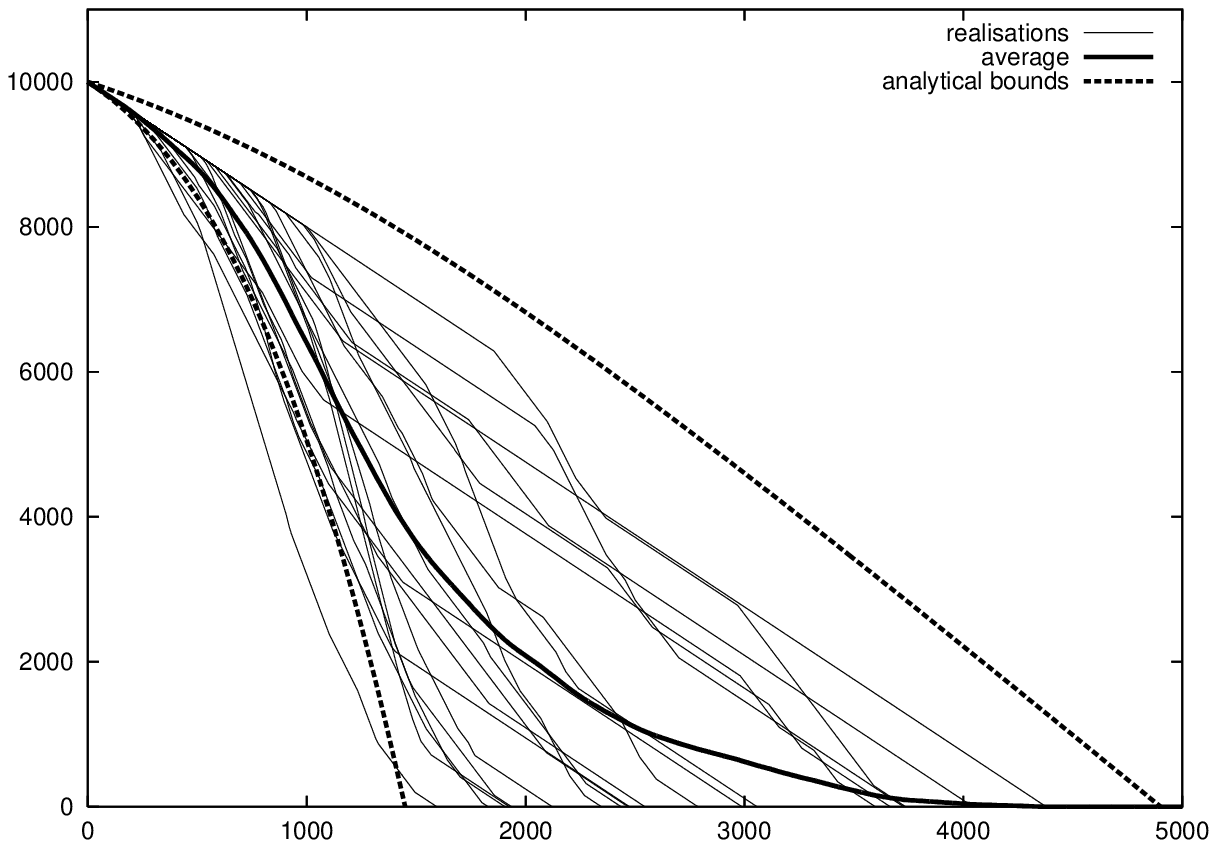,width=8.5cm,height=6.5cm}\!\!\!\!\!\!\!
            \epsfig{file=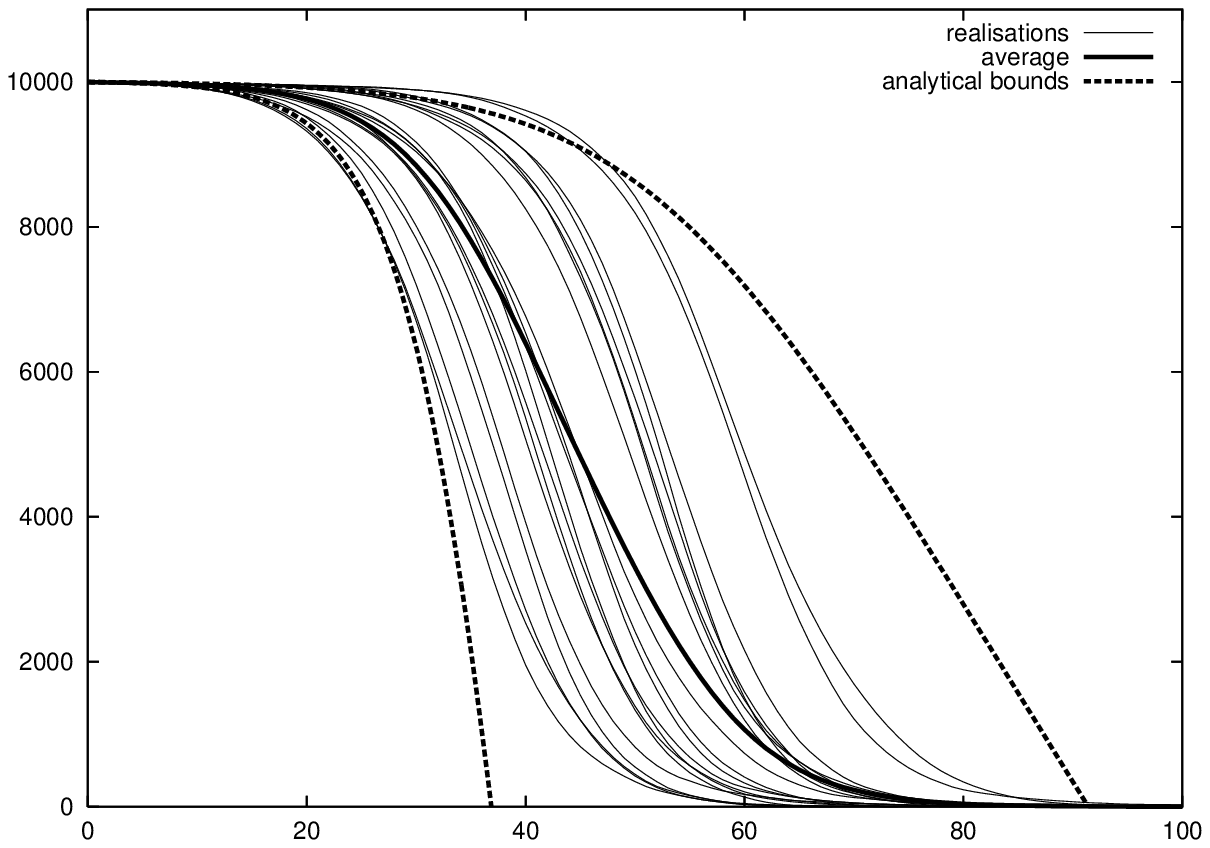,width=8.5cm,height=6.5cm}}
\centerline{\epsfig{file=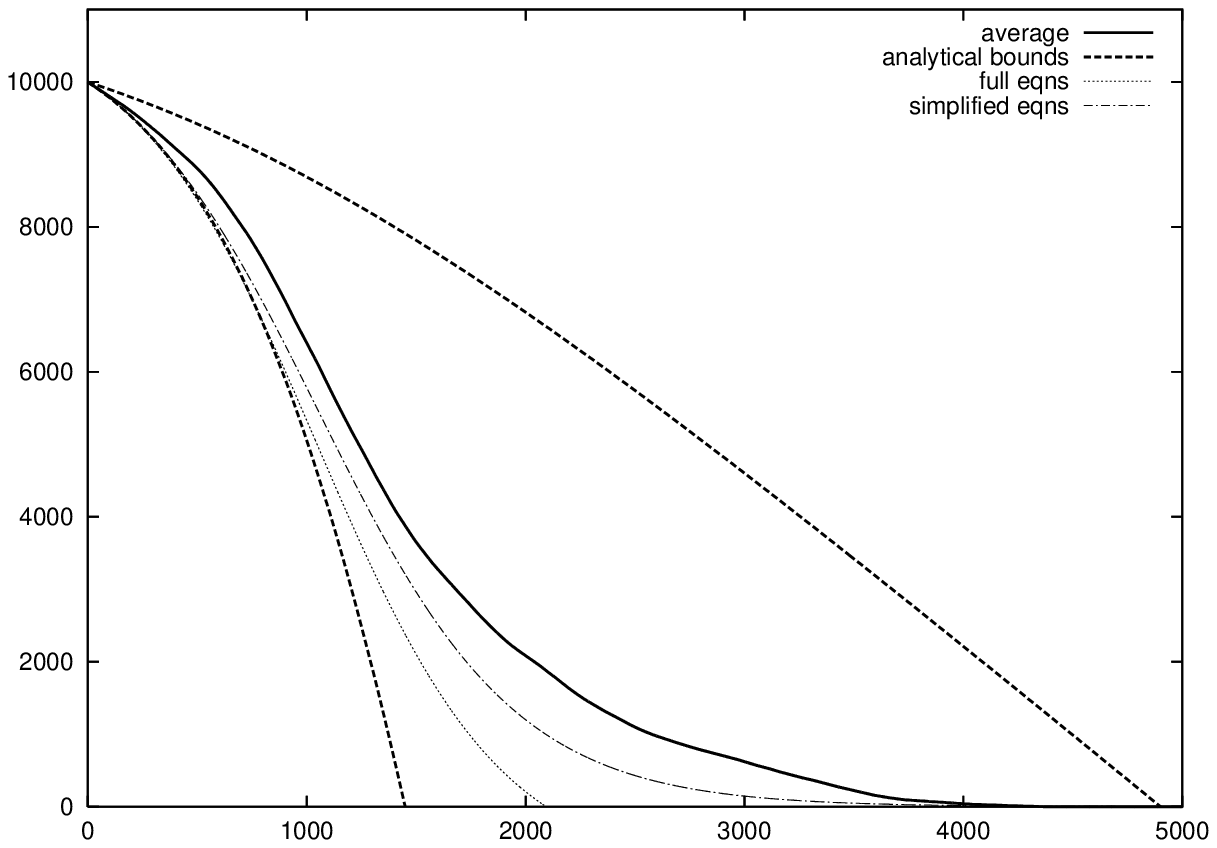,width=8.5cm,height=6.5cm}\!\!\!\!\!\!\!
            \epsfig{file=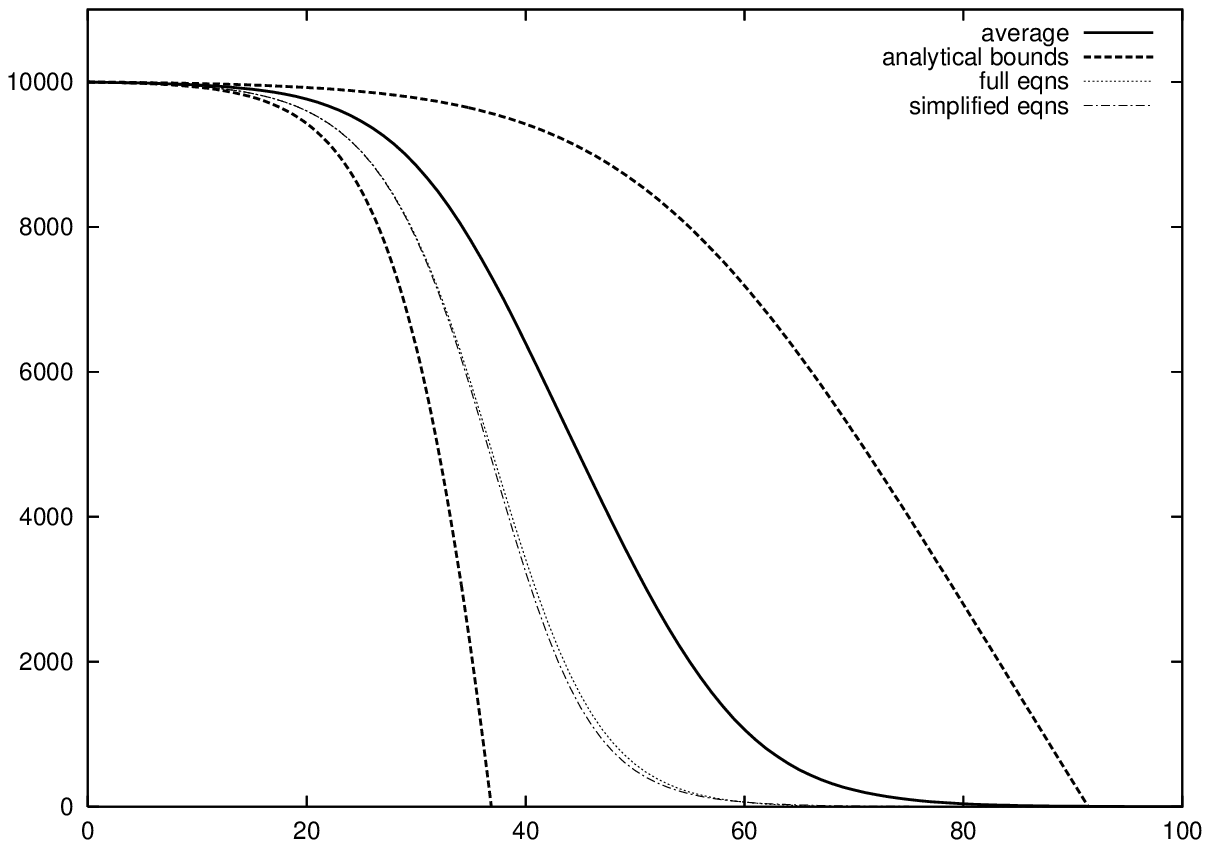,width=8.5cm,height=6.5cm}}
\caption{}
\label{diagfexamples}
\end{figure}

\begin{figure}[ht]
\centerline{\epsfig{file=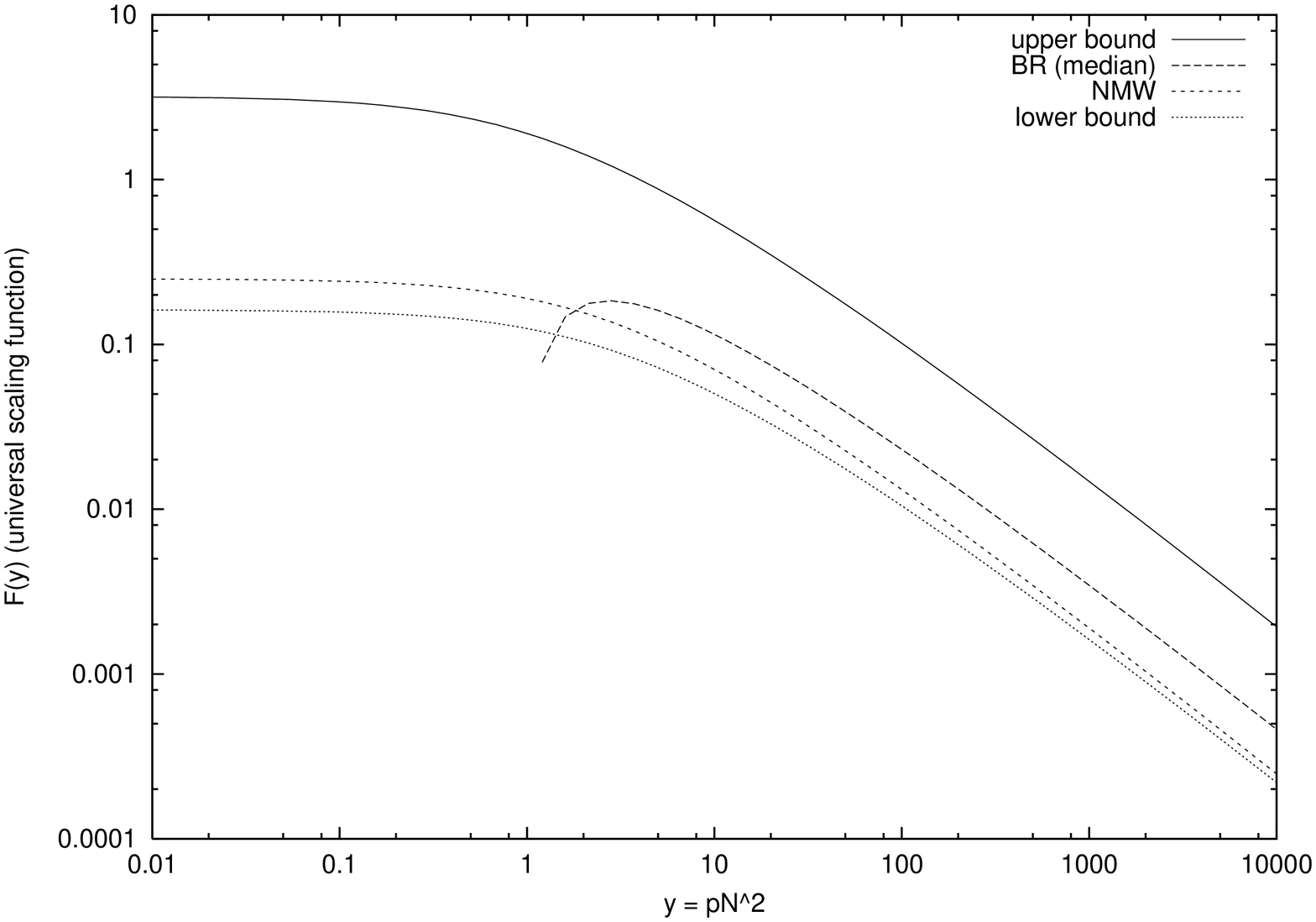,width=8cm,height=12cm,angle=270}}
\caption{}
\label{diagscalingfunction}
\end{figure}

\clearpage
\centerline{\large{\bf Figure Captions}}\vspace{0.5cm}
Caption Figure 1: The smallworld model for $k=1$. The number of nodes
is $N=30$. In this particular realisation we have inserted four
additional shortcuts. The unfilled nodes are the vertices which can be
reached by $\leq 3$ steps starting from node $x$, the step-number will
be denoted by $n$ in the following. The black nodes are the vertices
not reached after three steps, their cardinality being denoted by
$f(n)$ (free nodes).  This set consists of three connected subsets
which are separated by the subsets of nodes already reached. The
number of segments of free nodes is denoted by
$g(n)$ (gaps).\\[0.5cm]
Caption Figure 2: The function $f(n)$ for two examples. Left
side: $N=10^4, p=10^{-7}, \epsilon = \frac{3}{4}$; right side:
$N=10^4, p=10^{-5}, \epsilon = \frac{1}{4}$ (with $p$ normalized to
$N{-(1+\epsilon)}$ ). The upper diagrams show 20 realisations, their
mean value (averaged over fixed $n$) and the analytical bounds of
(\ref{ftildeeqn}). The lower diagrams show these bounds and the
averaged curve again, together with the numerical solutions of the
difference eqns (\ref{fulldiffeqg},\ref{fulldiffeqf}) and
(\ref{diffeqg}, \ref{diffeqf}). In both cases the averaged curve
exceeds the solution of the difference eqns (notice however the
perhaps surprisingly large standard deviation), but doesn't top the
upper analytical bounds.  The solutions of the full eqns
(\ref{fulldiffeqg}, \ref{fulldiffeqf}) only differ notable from the
simplified ones (\ref{diffeqg}, \ref{diffeqf}) in the left, nearly
linear case, the neglection of the additional combinatorial terms was
hence justified. Yet, what can't be seen in these diagrams is that, in
contrast to the full ones, the solution of the simplified eqns does
not sink to zero, and even crosses the upper boundary for large $n$.
This is however neither surprising nor important, as our boundary eqns
are only guaranteed to hold in the interval $f(n)\in[N/2,N]$, thus we
omit an additional logarithmical diagram.\\[0.5cm]
Caption Figure 3: Universal scaling functions

\end{document}